\documentstyle[11pt]{article}

\begin{document}

\setlength{\baselineskip}{0.8 cm}

\begin{center}
{\Large \bf Axial Currents in Electrodynamics\footnote{Work partially
supported by CNPq and FAPESP}}
\end{center}

\begin{center}
{\large \bf P.C.R. Cardoso de Melo$^{(1)}$, M.C. Nemes$^{(1)}$ and
Saulo C.S. Silva$^{(2)}$}
\end{center}

\begin{center}
$^{(1)}$Instituto de Ci\^encias Exatas, Universidade Federal de Minas
Gerais\\
CP702, 30000,
Belo Horizonte, MG, Brasil.
\end{center}

\begin{center}
$^{(2)}$Instituto de F\'{\i}sica, Universidade de S\~ao Paulo\\
CP20516, 01498, S\~ao Paulo, SP, Brasil.
\end{center}

\begin{abstract}
In the present work we argue that the usual assumption that magnetic
currents possess the vector structure characteristic of electric
currents may be the source of several difficulties in the theory of
magnetic monopoles. We propose an {\it axial} magnetic current instead
and show that such difficulties are solved. Charge quantization is shown to be
intimately connected with results of theories of discrete space time.
\vspace{0.5cm}

\end{abstract}

In 1931 Dirac proposes for the first time an electromagnetic theory
with magnetic monopoles$^{[1]}$. Such polemic subject has fascinated
many generations of physicists ever since. Its appeal is mainly connected
to  the
(up to now) unique possibility of explaining the quantization of the
electric charge.

Dirac's hypothesis, in spite of its undeniable
theoretical appeal, brings out some important difficulties.
Firstly, in Dirac's theory one is
faced with a symmetry problem: the terms responsible for the monopole in
the generalized Maxwell's equations violate their symmetry under space
and time reversal.  This new asymmetry remains up to
now an open problem. Secondly, Dirac's quantization condition implies in half
integer values for the electromagnetic field angular momentum. Finally, what
seems to be the main problem:
there has been, since Dirac's proposal, no conclusive experimental
evidence of magnetic monopoles.

In this work we propose a new magnetic current, namely an {\it axial
magnetic current} which presents the following differences as compared to
previously proposed ones, always of vector structure:

\begin{description}

\item $a$) the resulting theory preserves space and time inversion
invariance without having to resort to a pseudo-scalar magnetic charge;

\item $b$) besides the usual conservation of the vector electromagnetic
current, we have also the conservation of an axial current;

\item $c$) the charge quantization in the present theory leads to the
conclusion that velocities are constrained to rational values only. This result
already integrates theories of discrete space time;

\item $d$) as opposed to Dirac's monopole theory, the quantum of the field's
angular momentum is not necessarily half integer;

 \item $e$) we are able to reinterpret the current experimental results,
explaining their ``negative'' conclusions and to suggest very special
and (we hope) more favourable experimental conditions for the
observation of the magnetic monopole;

\end{description}

We start with the generalized definition of the electromagnetic field
tensor

\begin{equation}
F_{\mu\nu}=\partial_{\mu}A_{\nu}-\partial_{\nu}A_{\mu}
+\epsilon_{\mu\nu\alpha\beta}\partial^{\alpha}B^{\beta}
\end{equation}

\noindent where $B^{\mu}$ represents the new potential as defined in
ref [2]. Maxwell's equations for the fields $A^{\mu}$ and $B^{\mu}$ in
Lorenz's gauge $(\partial^{\mu}A_{\mu}=\partial^{\mu}B_{\mu}=0)$
become

\begin{eqnarray}
\Box A_{\mu}=j_{\mu}\\
\Box B_{\mu}=g_{\mu}
\end{eqnarray}

The quantity $F_{\mu\nu}$ in (1) is a tensor;
$\epsilon_{\mu\nu\alpha\beta}$ is a pseudo-tensor and therefore the
field $B_{\mu}$ must be a pseudo-vector or an {\it axial field}. From the
point of view of quantum theory the field $B_{\mu}$ represents
photon-like particles except for $P$, $T$ and $C$ parities. In other
words, {\it axial photons}. From
this it follows  that (3) is not invariant under
time and space reversal, unless $g^{\mu}$ is also a pseudo-vector. The
simplest possibility is the assumption that the
magnetic charge be a pseudo-scalar, if the current is of vectorial character.

We shall adopt here, however, a different hypothesis: {\it magnetic
monopoles are spin $1/2$ fermions, the magnetic charge $g$ is a true
scalar and the corresponding current is an axial vector current}.
Namely,

\begin{equation}
g_{\mu}=-g\bar{\psi}\gamma_{\mu}\gamma_{5}\psi
\end{equation}

Let us investigate the consequences of such current to the equations
which govern the
electromagnetic fields. The essential differences are given by the
equation

\begin{equation}
\partial^{\nu}F_{\nu\mu}^{\dagger}=g_\mu = -g\bar{\psi}\gamma_{\mu}
\gamma_{5}\psi
\end{equation}

\noindent
where $F_{\nu\mu}^{\dagger}$ corresponds to $F_{\nu\mu}$'s dual tensor. Since
 $F_{\nu\mu}^{\dagger}$ is antisymmetric one gets

\begin{equation}
\partial^\mu g_\mu =0
\end{equation}

\noindent
which means axial current conservation.

In terms of electric and magnetic fields $\vec{E}$ and $\vec{H}$ one
obtains

\begin{eqnarray}
\vec{\nabla}\cdot\vec{H} & = & -g\bar{\psi}\gamma_{0}\gamma_{5}\psi=
-g\psi^{\dagger}\gamma_{5}\psi\\
\vec{\nabla}\times\vec{E} & = & -\frac{\partial\vec{H}}{\partial t}
-g\bar{\psi}\vec{\gamma}\gamma_{5}\psi=
-\frac{\partial\vec{H}}{\partial t}
-g\psi^{\dagger}\vec{\alpha}\gamma_{5}\psi
\end{eqnarray}

For the sake of argument we now consider the nonrelativistic limit, in
which the monopole's velocity is small $v\ll 1$, and the magnetic
current in the classical limit. The second component of $\psi$, $\chi$,
is in this limit given by

\begin{equation}
\chi\simeq\frac{1}{2}\vec{\sigma}\cdot\vec{v}\phi\ll\phi
\end{equation}

\begin{equation}
\psi^{\dagger}\gamma_{5}\psi=\left(
\begin{array}{c c}
\phi^{\dagger} & \chi^{\dagger}\\
\end{array}
\right)\left(
\begin{array}{c}
-\chi\\
-\phi
\end{array}
\right)=
-\left(\phi^{\dagger}\chi+\chi^{\dagger}\phi\right)=
-\phi^{\dagger}\lambda\left|\vec{v}\right|\phi
\end{equation}

\begin{equation}
\psi^{\dagger}\vec{\alpha}\gamma_{5}\psi=\left(
\begin{array}{c c}
\phi^{\dagger} & \chi^{\dagger}\\
\end{array}
\right)\left(
\begin{array}{c}
-\vec{\sigma}\phi\\
-\vec{\sigma}\chi
\end{array}
\right)\simeq
-\phi^{\dagger}\vec{\sigma}\phi
\end{equation}

\noindent where $\lambda/2$ is the expectation value of the monopole's
helicity. Substituting the above expressions in (7) and (8),

\begin{eqnarray}
\vec{\nabla}\cdot\vec{H}=\rho_{m}\lambda\left|\vec{v}\right|\\
\vec{\nabla}\times\vec{E}=-\frac{\partial\vec{H}}{\partial t}+
\rho_{m}\vec{\sigma}
\end{eqnarray}

\noindent $\rho_{m}=\phi^{\dagger}\phi g$ stands for the magnetic
charge density, $\vec{\sigma}$ is the vector corresponding to the
expectation value of the spin operator $\hat{\vec{\sigma}}$, with

\begin{equation}
\rho_m \vec{\sigma} \equiv g \phi^{\dagger} \hat{\vec{\sigma}} \phi
\end{equation}

{}From equations (12) and (13) one sees that even if $g$ is large ($g=2\pi/e$,
$n=1$), the contribution of the current $g_\mu$ to $\vec{\nabla}\cdot \vec{H}$
 and $\vec{\nabla} \times \vec{E}$ may well be very small. This is basically
due to three effects: $\left| \vec{v} \right|$, $\rho_m$ or $\lambda$ may
independently be very small. On the light of such equations it is possible to
reinterpret the experiments involving accelerators or cosmic rays.

In the present scheme the effects due to magnetic monopoles should become most
conspicuous in experiments with polarized beams, at sufficiently high energies
and densities. Experiments with non-polarized beams are therefore not
conclusive. It is easy to check that
 the same conclusions can be reached as to the
magnitude of the effect of external fields on the monopoles. For the
same reasons they will also be negligible, unless under the very special
circumstances mentioned above: the Lorentz force density in this case
is

\begin{equation}
\vec{f}=\lambda\left|\vec{v}\right|\rho_{m}\vec{H}-\rho_{m}
\vec{\sigma}\times\vec{E}
\end{equation}

Let us now analyze the compatibility of the axial magnetic current with charge
quantization. We shall approach the problem in two different ways.

We first consider the gauge invariant wave function of a charged spin $1/2$
field in the presence of the electromagnetic field (see ref [2]),

\begin{eqnarray}
\Phi_{e}\left(x,P'\right)  & = & \Phi_{e}\left(x,P\right)
\exp\left[-\frac{\imath e}{2}\int_{S} F^{\mu\nu}d\sigma_{\mu\nu}
\right]
\end{eqnarray}

\noindent $S$ being any surface with contour $P'-P$. Due to the
arbitrariness of the surface $S$ we can write

\begin{equation}
\Phi_{e}\left(x,P\right)
\exp\left[-\frac{\imath e}{2}\int_{S} F^{\mu\nu}d\sigma_{\mu\nu}
\right]=
\Phi_{e}\left(x,P\right)
\exp\left[-\frac{\imath e}{2}\int_{S'} F^{\mu\nu}d\sigma_{\mu\nu}
\right]
\end{equation}

\noindent which leads to the condition

\begin{equation}
\exp\left[-\frac{\imath e}{2}\oint_{S-S'} F^{\mu\nu}d\sigma_{\mu\nu}
\right]=1
\end{equation}

\noindent or equivalently to

\begin{equation}
\exp\left[-\imath e\int_{V} \partial^{\nu}F_{\nu\mu}^{\dagger}dV^{\mu}
\right]=1
\end{equation}

\noindent where $V$ is the volume corresponding to the arbitrary
surface $S-S'$. We have

\begin{equation}
\partial^{\nu} F_{\nu\mu}^{\dagger}=g_{\mu}\neq 0
\end{equation}

\noindent and therefore

\begin{equation}
\exp\left[-\imath e\int_{V}g_{\mu}dV^{\mu}\right]=1
\end{equation}

Using our definition of $g_{\mu}$ (4), we get

\begin{equation}
Q_{V}=\int_{V}\left(-g\bar{\psi}\gamma_{\mu}\gamma_{5}\psi\right)
dV^{\mu}=\frac{2\pi n}{e}
\end{equation}

As $Q_{V}$ is a Lorentz scalar, we can perform the calculation in a
convenient reference frame. We choose a reference frame in which the monopole
velocity is constant and $v \ll 1$. For $\vec{\sigma}$ and $\vec{v}$ in the
same direction (e.g. $z$ direction), using the previous results for $\psi$ in
this limit we obtain

\begin{equation}
Q_v = g \left\{ -v \int \phi^{\dagger}\;\phi\;dx\;dy\;dz + \int
\phi^{\dagger}\;\phi\;dx\;dy\;dt\right\}=\frac{2\pi n}{e}
\end{equation}

Since $z=vt ,\;\;\;dt=dz/v$,

\begin{equation}
Q_v = g \left\{ -v + \frac{1}{v} \right\} \int
\phi^{\dagger}\;\phi\;dx\;dy\;dz= \frac{2 \pi n}{e}
\end{equation}

\noindent
which  gives

\begin{equation}
Q_v = g \left\{\frac{1}{v} - v\right\} \simeq \frac{g}{v} = \frac{2\pi n}{e}
\end{equation}

\noindent
or

\begin{equation}
\frac{eg}{2\pi v} = n
\end{equation}

The same result can alternatively be obtained semiclassically.
As a consequence of the space-like nature of our magnetic
current, we will not have radial fields for monopoles at rest (see
(12)). This seems to suggest that an electric charge would
not feel the action of such monopoles. In this case Goldhaber's
derivation of Dirac's condition$^{[3]}$ would not be valid. However,
this is not the case: a monopole at rest generates an electric field
with nonvanishing curl, (see (13))
which is a sufficient condition for the
validity of Goldhaber's approach to Dirac's charge quantization
condition. This electric field is given by ``Coulomb's law"

\begin{equation}
\vec{E}= \frac{g}{4\pi}\vec{\sigma}\times \frac{\vec{r}}{r^3}
\end{equation}

Consider the scattering of an electric charge by this  field$^{[4]}$. We assume
that the initial charge's velocity is along the direction of the vector
$\vec{\sigma}$. In such case, the variation of the charge's angular momentum
in this direction will be given by

\begin{equation}
\Delta L_{\sigma} = \frac{eg}{2\pi v}
\end{equation}

\noindent
independently of the impact parameter value. Using Bohr's quantization rule

\begin{equation}
\Delta L_{\sigma} = n
\end{equation}

\noindent
one gets, as before,

\begin{equation}
\frac{eg}{2\pi v} = n
\end{equation}

\noindent
where, now, $v$ is the charge's velocity.

Let us now proceed to the analysis of the conditions (26) and (30).
They will be fulfilled if one has simultaneously

\begin{equation}
\frac{eg}{2\pi}=m
\end{equation}

\begin{equation}
v=\frac{m}{n}
\end{equation}

The first equation is the celebrated Dirac's condition for charge quantization.
The second equation implies that the axial monopole's velocity (or electric
charge's velocity) must be a rational number. This result is already integrated
in discrete space time theories$^{[5],[6]}$, since its initial proposition by
 Yukawa.
It has recently been shown$^{[7]}$ that demanding rational values for the
 velocity
represents no contradiction at all with currently important theoretical
results,
such as Lorentz invariance, nor with experiments.

We last come to the question of the quantum of the field's angular momentum.
As discussed in ref $[4]$, the angular momentum of the field produced by the
pair charge-monopole, before and after scattering, is given by

\begin{equation}
L_{em}=\frac{eg}{4\pi}
\end{equation}

\noindent
In order to be compatible with Dirac's quantization condition, $L_{em}$ must
be quantized by half integer values! No such problem arises in the present
framework. It is simple to check that $L_{em}$ is zero before and after
scattering, since $\vec{\sigma}$ is held fixed in z-direction. Therefore one
is not confronted with the problem of having to impose a quantization
condition in this situation.

In conclusion we see that the introduction of a conserved axial magnetic
current, as proposed here, allows for a consistent parity conserving
electromagnetic interaction, without having to resort to a pseudoscalar
coupling constant. We are also able to solve the puzzle posed by the
quantization of the angular momentum of the field. Finally, the condition
of rational velocities obtained in the context of discrete space time
theories is shown to arise naturally in connection to charge quantization.
As to the observability of axial magnetic monopoles we show that it is
intimately connected to polarization conditions.

\end{document}